\documentclass[final,1p,times]{elsarticle} % do not change
\usepackage{graphicx} % do not change
\usepackage{amssymb} % do not change
\usepackage{amsthm} % do not change
\usepackage{lineno} % do not change

\journal{Nuclear Physics A} % do not change
\begin{document} % do not change

\begin{frontmatter} % do not change

%% QM09Author: please enter your  
%% Title, author and address info here; please do not use footnotes

% Your Title - please modify
\title{The Role of Color-Magnetic Monopoles in a Gluonic Plasma}

% Principle author, and co-authors - please modify
\author{Claudia Ratti}

% Address - please modify
% note that if you have authors from several institutions, we recommend
% labelling these [a], [b], [c] etc.
\address{Department of Physics \& Astronomy,
State University of New York at Stony Brook, % labe l [a]
Stony Brook, NY, 11794-3800, USA}

\begin{abstract} % do not change
%% Text of abstract goes here - please modify
The role of color-magnetic monopoles in a pure gauge plasma at high 
temperature $T>2T_c$ is considered.
In this temperature regime, monopoles can be considered heavy,
rare objects embedded into
matter consisting mostly of the usual ``electric" quasiparticles, 
quarks and gluons. 
The gluon-monopole scattering is found to hardly
influence thermodynamic quantities, yet it produces a
large transport cross section, significantly exceeding that
for pQCD gluon-gluon scattering up to quite high $T$. This mechanism
keeps viscosity small enough for hydrodynamics to work at LHC.

\end{abstract} % do not change

\end{frontmatter} % do not change

%% QM09: we keep linenumbers at least for initial version
%\linenumbers % do not change

%% start of main text - please modify. Below is a sub-set (single section) 
%% of an earlier proceedings that show how one can handle references 
%% and figures etc.
%%\section{}\label{}

\section{Introduction}
Creating and studying Quark-Gluon Plasma (QGP) in the laboratory
has been the goal of experiments at CERN SPS and recently
 at the Relativistic Heavy Ion Collider (RHIC)
facility in Brookhaven National Laboratory, soon to be
continued by the ALICE collaboration at the Large Hadron Collider (LHC).
  RHIC experiments have revealed 
robust collective phenomena in the form of radial and elliptic flows,
which turned out to be quite accurately described by near-ideal
 hydrodynamics. QGP thus seems to be the most perfect
liquid known, with the smallest viscosity-to-entropy 
ratio $\eta/s$.

Recently, the $electric-magnetic$ duality has been proposed, and used
to explain unusual properties of the QGP
\cite{Liao:2006ry}: in this so-called ``magnetic scenario",
the near-$T_c$ region is dominated by
 magnetic monopoles. 
 An important feature is 
 the opposite running of the electric coupling $e$ and the
magnetic one $g$, induced by the Dirac condition $e g=const$.
 As recently shown in \cite{Liao:2008jg}, this feature has been
dramatically confirmed
by the behavior of the lattice correlation functions 
\cite{D'Alessandro:2007su}, which indeed display 
monopole-monopole and antimonopole-monopole correlations
$increasing$ with $T$.
We consider the correlations observed in \cite{D'Alessandro:2007su}
  to be a decisive
confirmation of the existence of the
 long-distance magnetic Coulomb
field of the monopoles. It is natural to investigate the role played by these
objects in the QGP: 
we therefore address the issue of QGP 
transport properties in the ``magnetic scenario'' framework
\cite{Ratti:2008jz}.
We move  away from the phase transition region to higher temperatures,
where QGP is still dominated by the usual
 electric quasiparticles  -- quarks and gluons -- and the coupling is
 moderately small.
Our goal is to
study the interaction between electric and magnetic sectors.
Our main result is the explicit solution of the problem of quantum 
gluon-monopole scattering, from which we calculate the
 corresponding transport cross sections.
 \section{Quantum gluon-monopole scattering}
 The problem of quantum gluon-monopole scattering is solved in the pointlike
 monopole approximation. In this case, and for $j\neq 0$, the equations for the
 radial functions $T_{j\alpha}(\xi)$ reduce to
 generalized Bessel-like equations with noninteger index
 $j'=-\frac12\left[-1+
 \sqrt{(2j+1)^2-4n^2}\right]$, where $n=eg$ is the product of electric and magnetic
 couplings and $j$ is the total angular momentum quantum number\footnote{The pointlike monopole approximation is justified by the information about the
 monopole size that we obtain from the lattice, which
 indicate a monopole radius of $\sim 0.15$ fm \cite{Ilgenfritz:2007ua}.}:
 \begin{equation}
 T_{j\alpha}''(\xi)-\left[-\omega^2+1+\frac{j(j+1)-n^2}{\xi^2}\right]T_{j\alpha}(\xi)=0.
 \label{bessel}
 \end{equation}
 The index $\alpha$ runs from 0 to 9 and indicates all possible combinations of
 charge and spin polarization for gluons. After gauge fixing, three
 combinations turn out to be unphysical, and only six survive.
 The radial solution of the gluon-monopole scattering is easy; the complications
 reside in the angular functions. In fact, classically the gluon moves on the
 surface of a cone; the angular functions therefore are
 modified vector spherical harmonics that describe the conical motion in the classical
 limit of large angular momentum. 
 The scattering phase that we obtain from Eq.~(\ref{bessel}) is $\delta_{j'}=-j'\frac{\pi}{2}$,
 independent of energy.
 This feature is very important, since the contribution of this kind of scattering to thermodynamics
 is given by the Beth-Uhlenbeck formula
 \begin{equation} 
\delta M_m={T\over \pi} \sum_j (2j+1) \int dk {d\delta_j\over dk} f(k,T)
\label{eqn_BU}
\end{equation}
which vanishes identically for a constant scattering phase. Therefore, we find that
the gluon-monopole scattering does not contribute to thermodynamics. There is an
exception to this result, for $j=0$. In this case, the gluon can penetrate the monopole core and
form bound states. We don't discuss this case here, for all details we refer the reader
to Ref.~\cite{Ratti:2008jz}.
\section{Transport cross section: results and conclusions}
\begin{figure}
\begin{minipage}{.48\textwidth}
\parbox{6cm}{
\scalebox{.8}{
\includegraphics{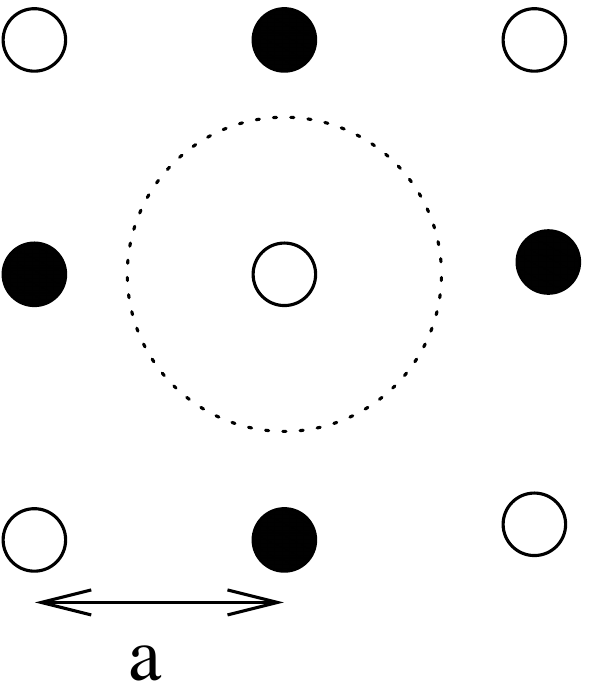}\\}}
\end{minipage}
%\hspace{.4cm}
\begin{minipage}{.48\textwidth}
\parbox{6cm}{
\scalebox{.65}{
\includegraphics{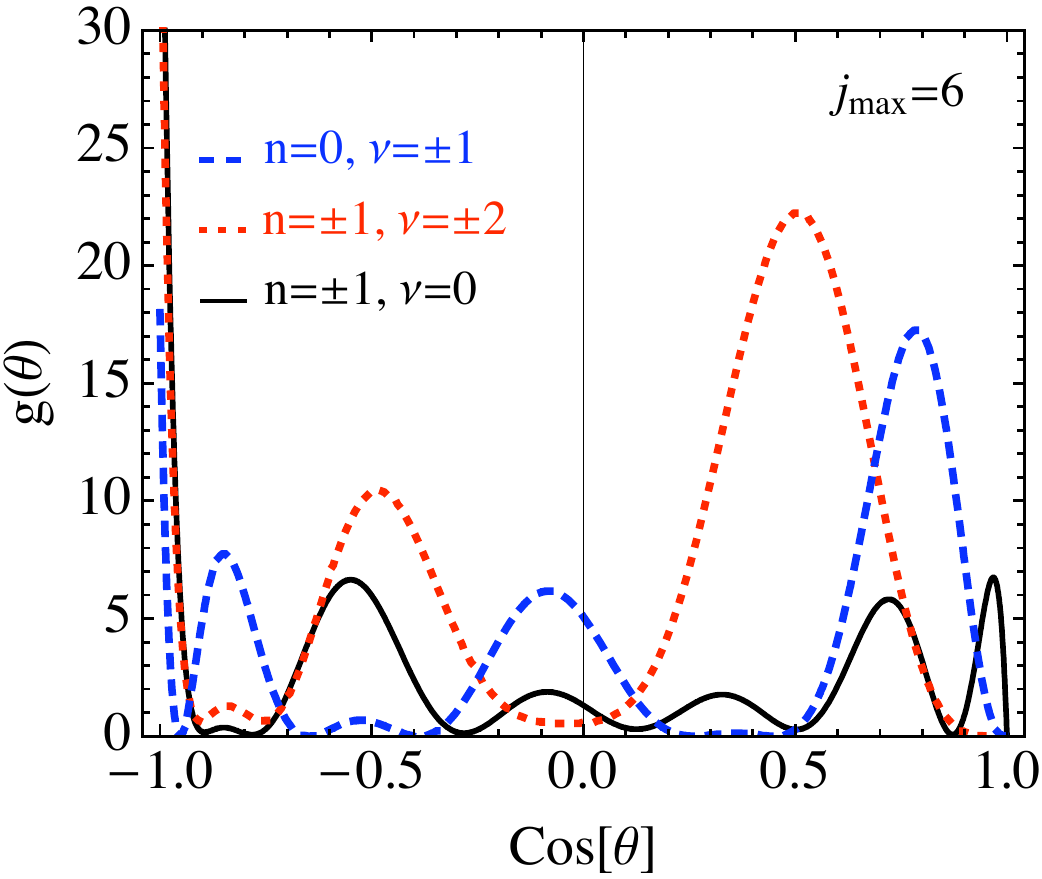}\\}}
\end{minipage}
\caption{Left: A charge scattering on a 2-dimensional array of 
correlated monopoles (open points) and antimonopoles (closed points).
The dotted circle indicates a region of impact parameters for
which scattering on a single monopole is a reasonable approximation.
Right: Integrand of the transport cross section 
$g(\theta)=(1-\cos(\theta))|f(\theta)|^2$
with only 6 lowest partial waves
included, for a gluon with $n=0,~\nu=\pm1$,
$n=\pm1,~\nu=0$ and $n=\pm1,~\nu=\pm2$. The strong peak backwards
is due to the presence of the cutoff $j_{max}$.}
\label{fig_monos_2d}
\end{figure}

The scattering amplitude $f(\theta)$ is given by the following formula:
\begin{equation}
2ik f(\theta)_{n,\nu}=\sum_{j=|\nu|}^{j_{max}} (2j+1)e^{i\pi (j'-j)} d^{(j)}_{\nu,-\nu}(\theta).
\label{scatteringamplitudev}
\end{equation}
where
$\nu=n+\sigma=\left(\vec{T}\cdot\hat{r}\right)+\left(\vec{S}\cdot\hat{r}\right)=-J_3.$
The sum over $j$ has an upper cutoff $j_{max}$: in matter there is a finite density of
 monopoles. A sketch of the setting,
assuming strong correlation of monopoles into a crystal-like
structure, is shown
 in the left panel of Fig. \ref{fig_monos_2d}. A
 ``sphere of influence of one monopole''(the dotted circle)
gives the maximal impact parameter
to be used. 
As a result,
 the impact parameter is limited from above by some $b_{max}$, which implies that
 only a finite number of partial waves should be included.
The
range of partial waves to be included in the scattering amplitude
can be estimated as $j_{max}=\langle p_x\rangle n_{m}^{-1/3}/2\sim aT\sim  1/e^2(T) \sim \log(T)$.
Since at asymptotically high $T$ the monopole
 density $n_{m}\sim (e^2 T)^3$ is small compared to
the density of quarks and gluons $\sim T^3$,
 $j_{max}$ asymptotically grows  logarithmically with $T$. So,
only in the academic limit
$T\rightarrow \infty$ one gets $j_{max}\rightarrow \infty$ 
and the usual free-space
scattering amplitudes calculated in \cite{Boulware:1976tv} 
where all partial
waves are recovered.
However, in reality
 we have to recalculate the scattering,
retaining only several lowest partial waves from the sum.
Taking the lattice results on the monopole density as a function
of the temperature \cite{D'Alessandro:2007su}, we estimate
$j_{max}\simeq 6$ in our temperature regime. 
This dramatically changes the angular distribution,
by strongly  depleting scattering at small angles and enhancing
scattering backwards. This is evident in the
right panel of Fig.~\ref{fig_monos_2d}, where
we show the angular distribution of the integrand
of the transport cross section $\sigma_t$:
\begin{equation}
(\sigma_t)_{n,\nu}=\int_{-1}^{1}d\cos\theta(1-\cos\theta)|f(\theta)_{n,\nu}|^2.
\end{equation}
The integrand exhibits a strong peak backwards, which would 
disappear in the absence of $j_{max}$.

\begin{figure}
\begin{minipage}{.48\textwidth}
\parbox{6cm}{
\scalebox{.6}{
\includegraphics{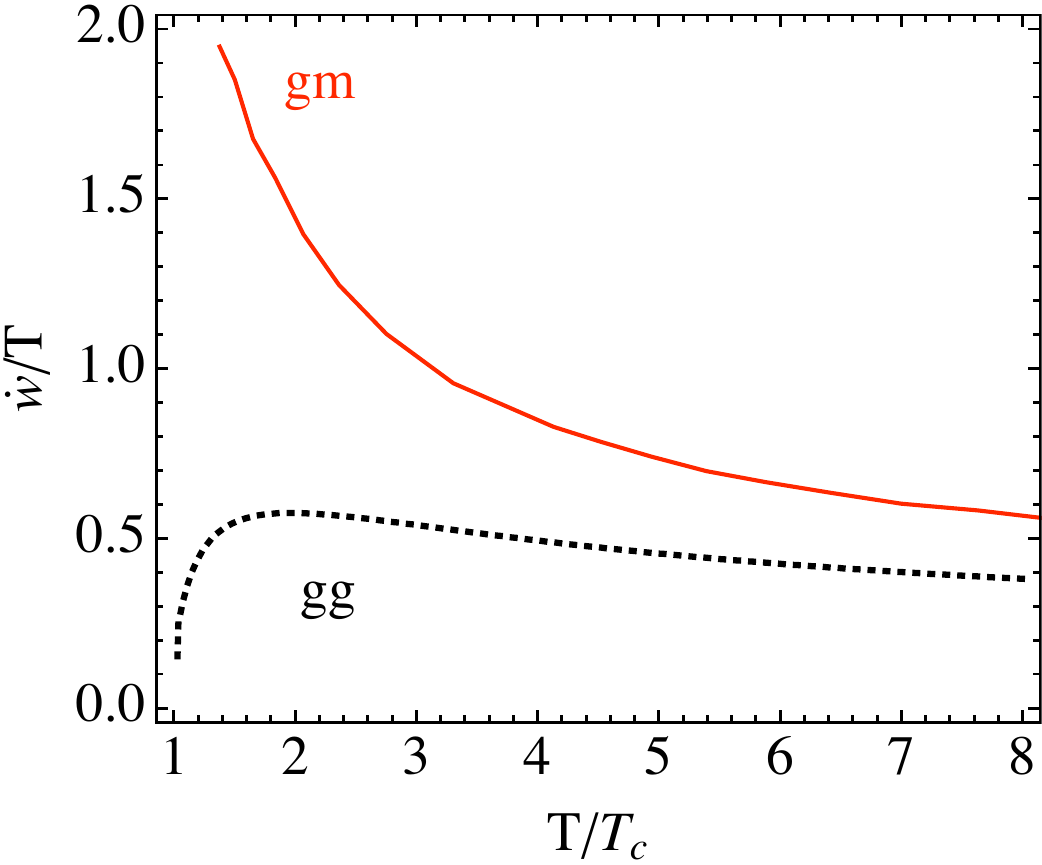}\\}}
\end{minipage}
\hspace{.4cm}
\begin{minipage}{.48\textwidth}
\parbox{6cm}{
\scalebox{.6}{
\includegraphics{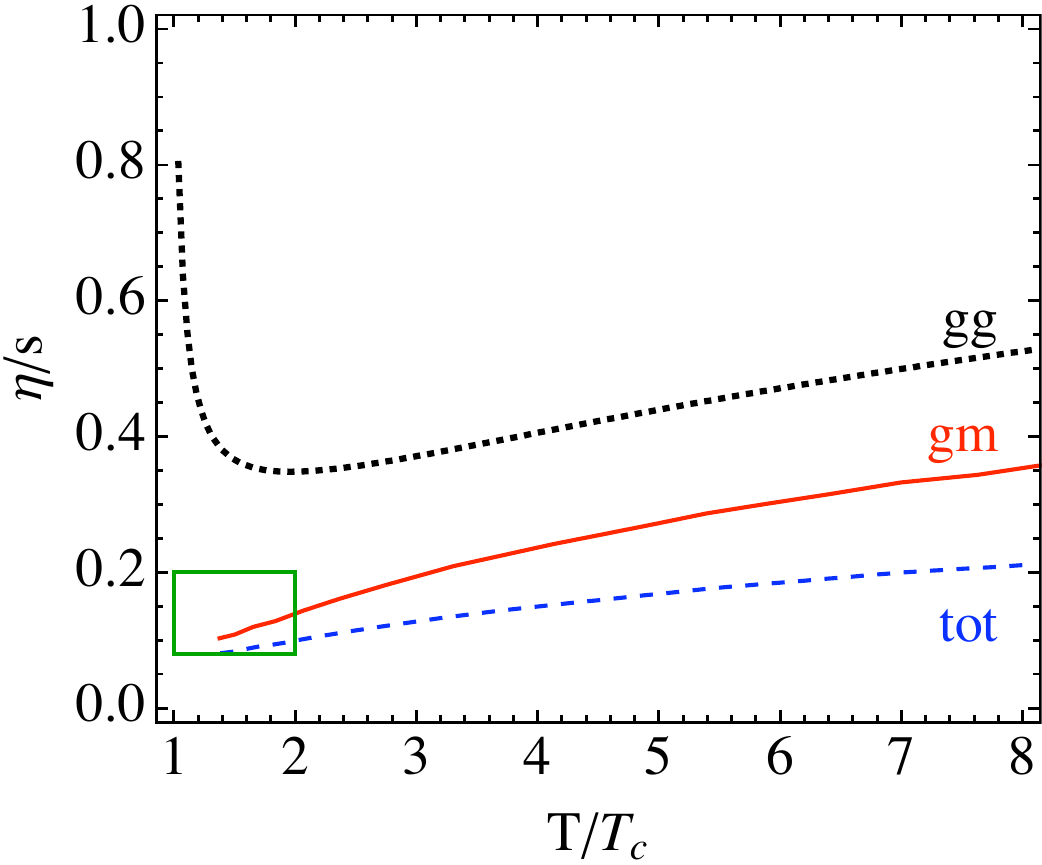}\\}}
\end{minipage}
\caption{Left panel: gluon-monopole and gluon-gluon scattering rate. Right panel: gluon-monopole
and gluon-gluon viscosity over entropy ratio, $\eta/s$. The blue, dashed curve is the total 
$\eta/s$, which is evaluated from the $gg$ and $gm$ contributions. The green box represents
the present estimate of $\eta/s$ in the RHIC temperature regime.}
\label{fig_viscosity}
\end{figure}

We now proceed to evaluate the scattering rate of gluons on monopoles:
\begin{equation}
\frac{\dot{w}_{gm}}{T}=\frac{\langle n_m (\sigma_t)_{gm}\rangle}{T}
\end{equation}
where the $\langle...\rangle$ indicates an average over the incoming gluon. $n_m$ is taken
from Ref. \cite{D'Alessandro:2007su} and is $n_m\simeq0.02$ GeV$^3$ in our temperature 
regime.
The gluon density has the following form:
\begin{eqnarray}
\!\!\!\!\!\!n_g(T)\!\!\!&=&\!\!\!\frac{8\pi}{(2\pi)^3}\int k^2dk\left[\frac{2}{\exp{(\beta\epsilon_k)}-1}
+\frac{2}{\exp{(\beta\epsilon_k)}\exp{(i\beta \mathcal{A}_{0}^{3})}-1}
+\frac{2}{\exp{(\beta\epsilon_k)}\exp{(-i\beta \mathcal{A}_{0}^{3})}-1}
\right.
\nonumber\\
\!\!\!&+&\!\!\!\left.
\frac{1}{\exp{(\beta\epsilon_k)}\exp{(2i\beta \mathcal{A}_{0}^{3})}-1}
+\frac{1}{\exp{(\beta\epsilon_k)}\exp{(-2i\beta \mathcal{A}_{0}^{3})}-1}\right]
=\frac{4\pi}{(2\pi)^3}\int k^2dk\rho_g(k,T)
\label{gluondensity}
\end{eqnarray}
where we have taken into account the suppression of electric particles due to the coupling
with the Polyakov loop (see for example \cite{PNJL}): $\mathcal{A}_{0}^{3}$
is a temporal background gauge field related to the Polyakov loop. In the average over the incoming gluon,
we have to take this suppression into account by integrating over $\vec{k}$
with the weight $\rho_g(k,T)$.
We show $\dot{w}_{gm}/T$ in the left panel of 
Fig. \ref{fig_viscosity} (the red, continuous line).
Also shown is the same quantity for the $gg$ scattering process (black, dotted line).
The approximate relation of the scattering rate to viscosity/entropy ratio is $(\eta/ s)\approx 
(T / 5 \dot w).$
We plot $\eta/s$ in the right panel of Fig. \ref{fig_viscosity}.
We observe a qualitative agreement between our results and the experimental
value for $\eta/s$ observed at RHIC, which is indicated in the right panel of Fig.
\ref{fig_viscosity} as a green box. Our present results
however deal with the purely gluonic sector of QCD only. For a more quantitative
and meaningful comparison with RHIC results, quarks need to be incorporated
in the analysis.
Our main finding is that 
the contribution of gluon-monopole scattering
is very important for transport properties. While
the monopole density may be small, the $gm$ scattering
amplitudes have $e^2g^2\sim O(1)$ coupling instead of small $e^4\ll1$.
 Furthermore,
in our setting (with a
limited number of partial waves $j<j_{max}$  included)
 there is an additional enhancement for large angle (or 
even backward) scattering.
It follows from this comparison of the gluon-monopole
curve with the gluon-gluon one that the former remains the leading
effect till very high $T$, although asymptotically it is expected to get subleading.
The maximal $T$  expected  at LHC does not exceed 4$T_c$, where the total
$\eta/s\sim .15$. This value is well in the region which would ensure
hydrodynamical radial and elliptic flows, although deviations from
ideal hydro would be larger than at RHIC (and measurable!).
\section*{Acknowledgments} % please check/modify, comment out or delete if not needed
I thank Jinfeng Liao and Massimo D'Elia for useful discussions. I also thank Professor Edward Shuryak, with whom the work presented in this talk has been done. The work is partially supported by the US-DOE grants DE-FG02-88ER40388 and DE-FG03-97ER4014.

 % do not change 
\end{document}